\documentclass[twocolumn,prl,10pt,amsmath,amssymb,nofootinbib,showpacs,superscriptaddress,floatfix]{revtex4-1}

\DeclareFontFamily{U}{rcjhbltx}{}
\DeclareFontShape{U}{rcjhbltx}{m}{n}{<->rcjhbltx}{}
\DeclareSymbolFont{hebrewletters}{U}{rcjhbltx}{m}{n}

\usepackage{graphicx,epsfig}
\usepackage{amsmath,amssymb,bbm}
\usepackage{color}
\usepackage[usenames,dvipsnames]{xcolor}
\usepackage[colorlinks=true,linkcolor=Red,citecolor=Green,linktoc=page]{hyperref}
\usepackage{multirow}

\usepackage[varg]{txfonts}
\usepackage{ulem}
\usepackage{fancyhdr}
\usepackage[caption=false]{subfig}
\usepackage{balance}
\usepackage{float}
\usepackage{flushend}

\DeclareFontFamily{U}{rcjhbltx}{}
\DeclareFontShape{U}{rcjhbltx}{m}{n}{<->rcjhbltx}{}
\DeclareSymbolFont{hebrewletters}{U}{rcjhbltx}{m}{n}

\usepackage[caption=false]{subfig}

\DeclareCaptionListOfFormat{nparens}{#1#2}
\usepackage[varg]{txfonts}
\usepackage{ulem}
\usepackage{fancyhdr}
\newcommand{\rs}{\rm\scriptscriptstyle}

\newcommand{\addresses}[1]{
\thispagestyle{fancy} \lfoot{\parbox{\textwidth}{ \vspace{0.3cm}
 \rule{\textwidth}{0.2pt}
\hspace{-0.2cm} \textsf{\scalefont{0.80} #1} \vspace{-0.2cm}
\begin{center}{\scalefont{0.87} \thepage}\end{center}}}
\cfoot{} }

\DeclareMathSymbol{\lamed}{\mathord}{hebrewletters}{108}

\begin{document}
\title{Direct probe of the interior of an electric pion in a Cooper pair superinsulator}

\author{M.\,C.\,Diamantini}
\affiliation{NiPS Laboratory, INFN and Dipartimento di Fisica e Geologia, University of Perugia, via A. Pascoli, I-06100 Perugia, Italy}
\author{S.\,V.\,Postolova}
\affiliation{A.\,V.\,Rzhanov Institute of Semiconductor Physics SB RAS, 13 Lavrentjev Avenue, Novosibirsk, 630090 Russia}
\affiliation{Institute for Physics of Microstructures RAS, GSP-105, Nizhny Novgorod 603950, Russia}
	\author{A.\,Yu.\,Mironov}
	\affiliation{A.\,V.\,Rzhanov Institute of Semiconductor Physics SB RAS, 13 Lavrentjev Avenue, Novosibirsk, 630090 Russia}
\affiliation{Novosibirsk State University, Pirogova str. 2, Novosibirsk 630090, Russia}
	\author{L.\,Gammaitoni}
\affiliation{NiPS Laboratory, INFN and Dipartimento di Fisica e Geologia, University of Perugia, via A. Pascoli, I-06100 Perugia, Italy}	
	 	\author{C.\,Strunk}
	 	\affiliation{Institute of Experimental and Applied Physics, University of Regensburg, D-93025 Regensburg, Germany}
	 	\author{C.\,A.\,Trugenberger}
	 	\affiliation{SwissScientific Technologies SA, rue du Rhone 59, CH-1204 Geneva, Switzerland}
	 \author{V.\,M.\,Vinokur}
\affiliation{Materials Science Division, Argonne National Laboratory, 9700 S. Cass Ave, Argonne, IL 60439, USA}
\affiliation{Consortium for Advanced Science and Engineering (CASE) University of Chicago, 5801 S Ellis Ave, Chicago, IL 60637, USA}
%
 
%

\begin{abstract}
The nature of hadrons is one of the most fundamental mysteries of physics. It is generally agreed that they are made of ``colored" quarks, which move nearly free at short scales but are confined inside hadrons by strong interactions at large distances. Because of confinement, quarks are never directly observable and, experimentally, their properties can be tested only indirectly, via high energy collisions. Here we show that superinsulating films realize a complete, one-color model system of hadron physics with Cooper pairs playing the role of quarks. We report measurements on highly controlled NbTiN films that provide a window into the interior of ``Cooper pair mesons" and present the first direct evidence of asymptotic freedom, ‘t Hooft’s dual superconductivity confinement mechanism, and magnetic monopoles.
\end{abstract}

\maketitle
\bigskip

\smallskip
\section*{Introduction}~~\\
Quark interactions are described by quantum chromodynamics (QCD), a non-Abelian gauge theory. A salient feature of QCD is asymptotic freedom\,\cite{gross1973,politzer}, the weakening of the interaction coupling strength at short distances (ultraviolet (UV) limit). At large distances (infrared
(IR) limit), the quarks are thought to be confined within hadrons,
which are physical observable excitations, by strings.
Quarks themselves cannot be extracted from hadrons and be seen in isolation.  The mechanism for the transition from weak quark
interactions in the UV regime to confinement and strings in the IR regime remains an open issue.

`t\,Hooft\,\cite{thooft1978} put forth an appealing confinement mechanism, the dual superconductivity where the condensate of magnetic monopoles\,\cite{goddard} constricts the chromoelectric field into thin flux tubes binding quarks
into mesons. He coined the term ``superinsulator", as opposite to superconductor, for the confined quark matter with infinite (chromo)-electric resistance. Polyakov showed that this confinement mechanism occurs also in Abelian gauge theories, provided they are compact and hence support topological excitations, magnetic monopoles, which are instantons in 2D and solitons in 3D\,\cite{polyakov_original, polyakov}. The fact that the compact QED in the confinement regime maps onto a confining string theory\,\cite{conf, dqt} makes it a perfect model for 't Hooft's dual superconductivity mechanism. 

Remarkably, Abelian confinement emerged in a condensed matter realization of superinsulators, predicted first for Josephson junction arrays (JJA) in\,\cite{Diamantini1996} and rediscovered in\,\cite{vinokur2008superinsulator} in films experiencing the superconductor-insulator transition (SIT). These electric superinsulators constitute a new state of matter with infinite resistance at finite temperatures due to electric strings binding Cooper pairs into ``electric mesons"\,\cite{dtv1}. Transport measurements revealed superinsulation in  titanium nitride (TiN) films\,\cite{Baturina2007,vinokur2008superinsulator}, niobium titanium nitride (NbTiN) films\,\cite{Mironov2018}, and, albeit under a different name, InO films\,\cite{Shahar2005,Shahar}. The long-distance electromagnetic response of superinsulators is exactly Polyakov's compact QED\,\cite{dtv1} with the effective coupling constant 
\begin{equation}
e_{\rm eff}^2 = 4\alpha f(\kappa) /\varepsilon g\ .
\label{coupling}
\end{equation}
Here $\alpha=e^2/(\hbar c)$ is the fine structure constant, $\kappa=\lambda/\xi$ is the Ginzburg-Landau parameter of the superconducting material, with $\lambda$ its London penetration depth and $\xi$ its coherence length and
$\varepsilon$ is the dielectric permittivity of the normal insulating state. Finally, $g$ is the tuning parameter driving the system across the SIT, so that $g\simeq 1$ near the transition. The function $f(\kappa)$ is smooth and is ${\cal O}(1/\kappa)$ for $\kappa\gg 1$\,\cite{dtv1}. Pure gauge compact QED in 2D is not renormalizable. However, coupling the action to dynamical matter results in a non-trivial fixed point\,\cite{kleinert1, kleinert2}. The same occurs in our case: compact QED is induced by an underlying matter dynamics from which it inherits the corresponding fixed point structure encoded in $g$. One can show\,\cite{bose, roman} that $g$, and hence also the effective coupling $e_{\rm eff}^2$, have a Berezinskii-Kosterlitz-Thouless (BKT) (see\,\cite{Minnhagen} for a review) infrared (IR) fixed point at the critical value $g_c$, at which the string tension diverges. The tension, and thus the interaction strength, flow to smaller values in the UV limit and, as a result, the induced compact ${\rm QED}_2$ becomes asymptotically free, albeit this applies near the confining IR fixed point, instead of near the UV-free fixed point, as in QCD. 

Confinement by strong interactions prevents a direct view on quarks despite that they move nearly free at small scales. Since electric Cooper pair mesons are generated by much weaker Coulomb interactions, they have a macroscopic dimension and are accessible to direct experimental study. Here we investigate superinsulators, which allow for a direct observation of the interior of electric mesons made of Cooper pairs by standard transport measurements. We reveal the transition from the confined to the asymptotic free Cooper pair motion upon decreasing the distance between electrodes, realizing the observation spatial scale. Using the compact QED mapping we demonstrate the electric Meissner effect and calculate the $I$-$V$-characteristics in the confinement regime. Comparing our experimental results with theoretical predictions yields the string tension and constitutes the first ever ``look inside a meson," directly confirming asymptotic freedom and 't Hoofts confinement mechanism by magnetic monopoles. 

\section*{Results}
\subsection{Strings and the Meissner and mixed states of a superinsulator}
A quantitative theory of the response of a superinsulator to a dc electric field rests on the fact that, in a superinsulator the fundamental excitations are electric strings with linear tension\,$\sigma$\,\cite{dtv1}.  The electric Meissner state is obtained from the dual analogue of the London equations\,\cite{qt,conf,dqt}, describing the electrodynamics of strings connecting charged particles. Deferring the technical derivation of the Meissner state and its vanishing static electric permittivity, $\varepsilon_{\rs SI}=0$, to the end of the paper, we note here that
the strings can be either closed, describing pure gauge excitations (the analogues of glueballs in QCD)\,\cite{caselle} or open, representing Cooper pairs -- anti-Cooper pair dipoles (the analogues of mesons in QCD). The presence of such Cooper pair dipoles follows from either self-induced or imposed electronic granular structure of a system supporting superinsulation. When a Cooper pair tunnels from one granule to another it leaves behind a +2e charge excess representing a Cooper pair ``hole". This picture is fully supported by the experimental observation of the charge BKT transition in such materials\,\cite{Mironov2018}. The energy to create pure gauge excitations is $\Delta_{\rs G}=mv^2$, with $m$ being the gauge-field mass and $v$$=$$c/(\varepsilon \mu)^{1/2}$ ($\mu \approx 1$ is the magnetic permeability) being the light velocity in the medium. Open strings have the typical length $d_{\rm s}= {\cal O}(\sqrt{\hbar v/\sigma})$ and a gap of the order ${\cal O}(\sqrt{\hbar v\sigma})$. The width of the strings is defined by the screening length $\lambda_{\rm el} = \hbar /mv$\,\cite{caselle}. Both $\Delta_{\rs G}$ and the string tension $\sigma$ are expressed via the ultraviolet (UV) cutoff $\Lambda_0 = \hbar v/r_0$, where $r_0\simeq\xi$ in films and is of order of the plaquette size in JJA, as functions of the effective coupling (\ref{coupling}) and $v$\,\cite{kogan}, 
\begin{equation}
m \propto {\hbar \over v r_0} {1\over e_{\rm eff}^2} {\rm exp} \left(- {\pi \over 4 e_{\rm eff}^2} \right) \ ,\,\,\,\,\sigma =  { e_{\rm eff}^2 v^2 \over 4 r_0} m \propto {\hbar v\over  r_0^2} {\rm exp} \left(- {\pi \over 4 e_{\rm eff}^2} \right) \ .
\label{mass}
\end{equation}

For samples with linear dimension $L \gg d_s$, small electric fields below a critical value $E_{\rm c1}$ are sufficient only to excite isolated strings of typical length $d_s$ much smaller than the sample size. In this regime, the applied electric field does not penetrate the sample, only neutral ``pion-like" dipole excitations made of a Cooper pair and a Cooper ``hole" can be created. This is the Meissner state of the superinsulator. 
When the applied electric field reaches the critical value $E_{\rm c1}$, enough electric pions can be created such that a chain of them reaches from one end of the sample to the other. At this point, a single electric flux tube can traverse the sample end-to-end and the Meissner state is destroyed. For $E > E_{\rm c1}$ electric fields penetrate the superinsulator in form of flux tubes of typical width $\lambda_{\rm el}$ and the mixed state of the superinsulator sets in. This is the dual state of the Abrikosov lattice in superconductors. Finally, at the critical field $E_{\mathrm{c2}}$, superinsulation breaks down.
\\
\begin{figure*}[!t]
	\center
	\includegraphics [width=14cm] {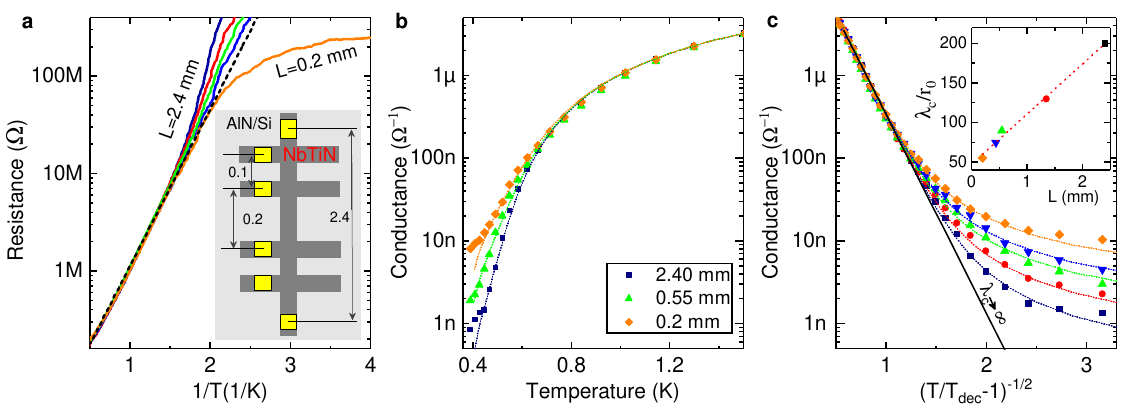}
	\caption{  \textbf{Insulating NbTiN film's  sheet resistance evolution with bridge length.}
		(\textbf{a}) The Arrhenius plot of sheet resistance $R_{\Box}$ vs. reversed temperature $1/T$ for fridges of various length $L$. Dashed line shows $R \propto \exp (1/T)$. Inset: Experimental setup. The Si substrate with AlN buffer layer is shown with light gray and Hall bridge of NbTiN is dark grey. The square gold contacts are given in yellow. All lateral sizes are given in millimeters.  The bridges lengths (i.e. distances between measuring electrodes) shown in the legend in panel (\textbf{b}). (\textbf{b}) Same data as in (\textbf{a}) but replotted as conductance $G=1/R_{\Box}$ vs. $T$ in log-line scale. A few curves are omitted to avoid overcrowding. Dotted lines are fits using a two dimensional Coulomb gas model\,\cite{Minnhagen} that generalizes the Berezinskii-Kosterlitz-Thouless (BKT) formula for conductance $G\propto\exp [-(T/T_{\mathrm{dec}}-1)^{1/2}]$ by incorporating a self-consistent solution to the effects of electrostatic screening\,\cite{Mironov2018}, where the screening length $\lambda_{\mathrm c}$ and $T_{\mathrm{dec}}$ enter as fitting parameters. For all bridges we find $T_{dec} \approx 400$\,mK.
		(\textbf{c}) Same data as in (\textbf{b}) but for temperature renormalized as  $(T/T_{\mathrm{dec}}-1)^{1/2}$. Solid line corresponds to the case of an infinite electrostatic screening length $\lambda_{\mathrm c} \rightarrow \infty$.
		Inset:  Screening length as function  of the bridge length $L$. Symbols correspond to BKT-fits of $G(T)$ (dotted lines in (\textbf{b}) and (\textbf{c})) and dashed line is the eye guide. The error bars are obtained from variations of parameters leaving self-consistent solutions of electrostatic equations of\,\cite{Mironov2018} unchanged and are much less than the size of symbols.}
	\label{FigPrinciple}
\end{figure*}

%

\subsection{Current-voltage characteristics}
We start the derivation of the $I(V)$ response by writing down the
compact QED potential for the interaction between Cooper pairs
\begin{equation}
U(r) = \sigma(T) r -{c\hbar \pi\over 24 r} + a \left[ {\rm ln} \left( {\lambda_{\rm el} \over r_0}\right)  - K_0 \left( {r\over \lambda_{\rm el}} \right) \right] \ ,
\label{potential}
\end{equation}
where the second term is the so-called L\"uscher term\,\cite{luscher}. The third term is the screened 2D Coulomb potential ($K_0$ is the MacDonald function) that reduces to  $a \ln (r/r_0)$ at $r$\,$\ll$\,$\lambda_{\rm el}$ and decays exponentially at $r$\,$\gg$\,$\lambda_{\rm el}$. At $r$$>$
$r_0$, the L\"uscher term is  negligible, so that $U(r_0)$\,$\simeq$\,$0$. Near the SIT, the strength of the Coulomb potential becomes\,\cite{dtv1}
\begin{equation}
a=(4e^2/2\pi \varepsilon_0 \varepsilon d) (f(\kappa)/g) \ .
\label{a}
\end{equation}
In the samples with sizes $\lambda_{\rm el} < L < d_s$, Cooper pairs feel neither the string tension nor the Coulomb interaction screened on the scale $\lambda_{\rm el}$. As we show below this is the asymptotically free regime. 

To relate superinsulating parameters to observable quantities, note that strings of length $d_{\mathrm s}$ have energy $\sqrt{\hbar v \sigma }$. Hence the energy to form a string chain spanning the entire system is $2eV_{\rm c1} = \sigma L$.  Accordingly, one can introduce the lower critical electric field for first string penetration the superinsulating sample
\begin{equation}
E_{\rm c1} = {\sigma \over 2e}\ ,
\label{lowercrit}
\end{equation}
which corresponds to the lower critical field $B_{\mathrm{c1}}$ in superconductors. For $E<E_{\rm c1}$ the Meissner state of superinsulators is realized: electric fields are completely expelled from the superinsulator by the electric dual of the Meissner effect. For $E > E_{\rm c1}$, instead, the mixed state of a superinsulator forms. Next, we introduce the deconfinement temperature $T_{\rm dec}$ that marks the transition between {\it linearly} bound charges  and unbound charges at $T>T_{\mathrm{dec}}$. This transition belongs in the BKT universality class and occurs via instanton condensation\,\cite{polyakov}. According to\,\cite{zarembo}
\begin{equation}
k_{\rs B} T_{\rm dec} = 8 a=  8 {f(\kappa)\over g} {\cal E}_{\rs C} \,
\label{aTdec}
\end{equation}
where ${\cal E}_{\rs C}=4e^2/2\pi \varepsilon \varepsilon_0 r_0$ is the characteristic bare Coulomb energy of the Cooper pair and  $a$ is defined by Eq.\,(\ref{a}).

We calculate the current as $I \propto 2en_{\rm f} V$, where $n_{\rm f}$ is the equlibrium density of free charges. For the external field $E_{\mathrm{ext}}$$<$$E_{\mathrm{c1}}$$\equiv$$\sigma/2e$ the maximum of the potential lies always at the distance $L$ corresponding to the sample size and, thus, the current is simply proportional to the number of charges activated over the barrier $(\sigma -2eE_{\rm ext} ) L $,
\begin{equation}
I \propto V{\exp} \left(- {\sigma(T)L \over k_{\rs B} T} \right)\,\, , \quad V < V_{\rm c1} = {\sigma(T)L\over 2e}\ ,
\label{exp}
\end{equation}
which, in the thermodynamic limit $L\to\infty$, implies an infinite resistance.

For $E_{\mathrm{c1}}$\,$<$\,$E_{\mathrm{ext}}$, the potential has a maximum at $r=r^*$ determined by the equation $K_1\left( r^*/\lambda_{\rm el} \right) = \left( 2e E_{\rm ext} -\sigma \right) \lambda_{\rm el}/a$. Two distinct regimes become possible. The first is realized in small samples such that
\begin{equation}
K_1\left( L/\lambda_{\rm el} \right) >  {4\lambda_{\rm el} \over d_s} {\Delta V\over V_{c1}} \ ,
\label{small}
\end{equation}
where $\Delta V$$=$$V-V_{\rm c1}$, and we have used $d_s = \hbar v/k_B T_{\rm dec}$. Then the potential for $\lambda_{\rm el}$\,$<$\,$r$\,$<$\,$L$ is essentially flat, implying the asymptotic free regime, where charges effectively do not interact, and we expect thus a metallic saturation of the resistance at the lowest temperatures. The ratio $d_s/\lambda_{\rm el}>1$  but not typically extremely large\,\cite{dtv1}. Also, the function $K_1(x)$\,$\simeq$\,$\exp(-x)$/$\sqrt{x}$ at $x$\,$\gg$\,$1$. Thus, the typical sample size for which  this metallic behavior emerges is $O(d_s)$, although it can become larger if measurements are taken just above $V_{c1}$.

In the limit opposite to (\ref{small}), the total energy $U_{\mathrm{E_{\mathrm{ext}}}}$ of the charge-anticharge pair following from (\ref{potential}) is
\begin{equation}
U_{\mathrm{E_{\mathrm{ext}}}}=a\ln(r/r_0)-Fr,
\label{potential1}
\end{equation}
where $F$\,$=$\,$2eE_{\mathrm{ext}}$\,$-$\,$\sigma$ is the effective force pulling the charge-anticharge pair apart. The saddle point\,$r^*$\,of this potential, controlling the activated current, is $r^*$\,$=$\,$a/F$, so that the energy barrier is $U^*$\,$\equiv$\,$U_{\mathrm{E_{\mathrm{ext}}}}(r^*)$\,$=$\,$a\left[ \ln(a/Fr_0)-1\right]$. In equilibrium, the ionization rate ${\cal R}$\,$\propto$\,$\exp(-U^*/k_{\rs B}T)$ and the recombination rate, ${\cal R}_{\mathrm r}$, of the\,$\pm$\,charges, are equal. Since ${\cal R}_{\mathrm r}$\,$\propto$\,${n_+}n_-$\,$=$\,$n_{\mathrm f}^2$, then with logarithmic accuracy $n_{\mathrm f}$\,$=$\,$\sqrt{{\cal R}}$\,$\propto$\,$\exp(-U^*/2k_{\rs B}T)$\,$\propto$\,$(Fr_0/a)^{a/(2k_{\rs B}T)}$, and Eq.\,(\ref{aTdec}) yields
\begin{eqnarray}
&&I\propto(V-\sigma L/2e)^{1+{T_{\mathrm{dec}}/16T}}\nonumber \\
&&V_{\mathrm{c1}}<V<V_{\mathrm{c2}}\equiv \frac{L}{2e}\left[\sigma(T)+\frac{k_{\rs B}T_{\mathrm{dec}}}{2.718r_0}\right]\,,
\label{VvsL}
\end{eqnarray}
where the upper critical voltage $V=V_{\mathrm{c2}}\equiv V_{c2}$ is the voltage where the energy barrier $U^*$ vanishes, the $I$-$V$ curve experiences a jump and the system switches from superinsulation into a normal insulating state. In the field interval such that $V_{\mathrm{c1}}<V<V_{\mathrm{c2}}$,
the superinsulator is in the mixed state, where an ensemble of electric strings penetrates the system. This is the analogue to the mixed, or Abrikosov state in superconductors.
\\

\begin{figure*}[!t]
	\center
	\includegraphics [width=14cm] {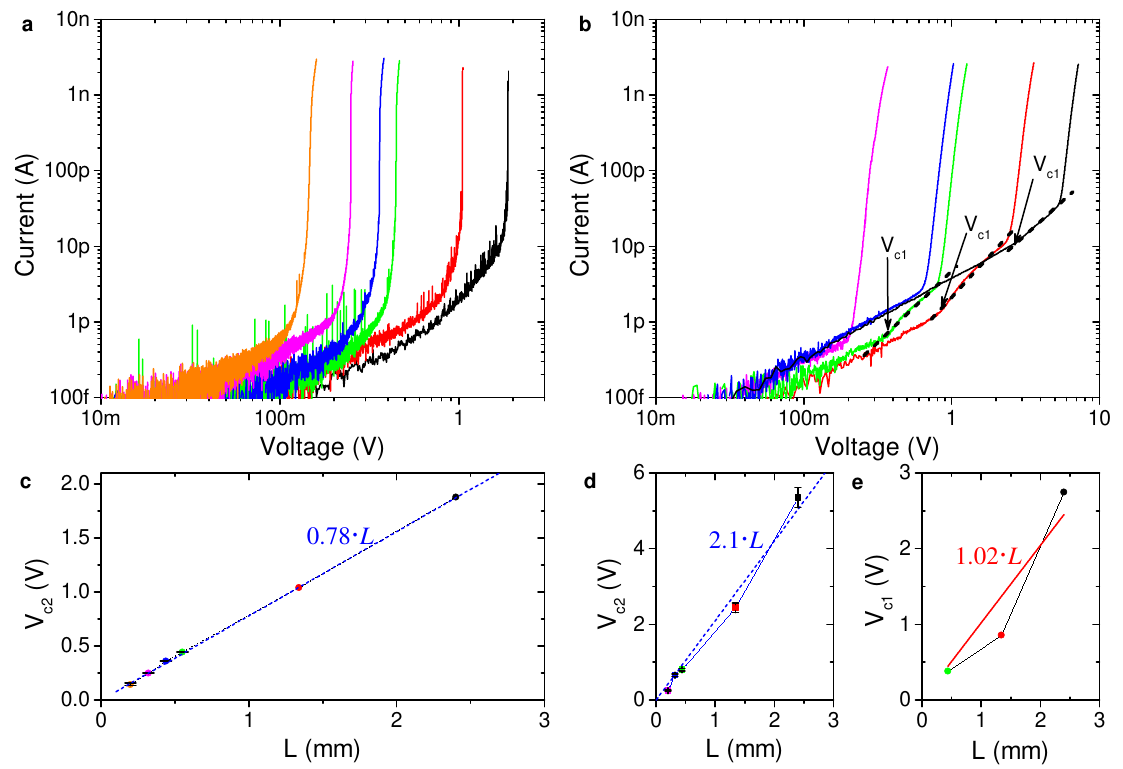}
	\caption{ \textbf{Threshold voltage of bridges of different length.}
		The $I$-$V$ curves of two different NbTiN samples with (\textbf{a}) $R_{\rs\square}(T$$=$2\,K$)=0.2$\,M$\Omega$ and (\textbf{b}) $R_{\rs\square}(T$$=$2\,K$)=2$\,M$\Omega$ taken at the same temperature $T=50$\,mK. Different colors correspond to different bridge's length $L$ (distance between electrodes). (\textbf{c}) Dependence of the threshold voltage $V_\mathrm{c2}$ on the bridge's length for 0.2\,M$\Omega$ sample. The dashed line depicts $V_{c2} = 0.78 \cdot L$ dependence. The error bars are defined as the width of the jump and are less than the symbol sizes.  (\textbf{d})  Threshold voltage $V_\mathrm{c2}$ for the 2\,M$\Omega$ sample; $V_\mathrm{c2} = 2.1\cdot L$ (dashed line). Error bars are given by the width of the jump. (\textbf{e}) The dependence of the kink voltage marked by arrows in main panel, which we associate with the critical voltage $V_{\mathrm{c1}}$ upon the bridge length. The error bars are determined from the errors in the slopes of the $I$-$V$ curves and are much less than the symbol sizes. }
	\label{FigScurve}
\end{figure*}

\subsection{Experiment}
The measurements are taken on a polycrystalline 10\,nm thin NbTiN film obtained by atomic layer deposition (ALD)\,\cite{Mironov2018} at temperature 350\,$^{\circ}$\,C. The temperature dependencies of the resistances are measured on a sample patterned by photolithography into the Hall bar with the width $50\,\mu$m, see inset in Fig.\,\ref{FigPrinciple}a. The chosen geometry enables measurements on bridges of different lengths. The experiment is carried out in a $^{3}$He/$^{4}$He dilution refrigerator. The resistance is measured by two-terminal circuit under the low-frequency,  $f \sim 1$\,Hz, ac voltage,  $V \sim 100\,\mu$V, in the linear regime as verified by the direct measurements of the films' $I$-$V$s.

The Arrhenius plots of the resistances per square $R_{\rs{\square}}(T)$ versus inverse temperature $1/T$ for bridges of different lengths of the same NbTiN film are shown in Fig.\,\ref{FigPrinciple}a.
At $T>0.8$\,K, $R_{\rs \square}(T)$ all curves collapse on top of each other. For temperatures below $T_{\rm dec} = 400$\,mK the long bridges show the hyperactivated behaviours characteristic of superinsulators\,\cite{Baturina2007,vinokur2008superinsulator}. The smallest bridge 0.2\,mm short, however, clearly shows a metallic-like saturation at very low temperatures. Using the measured critical temperature of 400\,mK,  the known dielectric constant $\varepsilon = 800$, and the coherence length $r_0 = \xi = 10$\,nm in NbTiN\,\cite{Mironov2018}, one can estimate the string size as $d_s \simeq \hbar v/ k_B T_{\rm dec}$, obtaining $d_s \simeq 0.13$\,mm in a remarkable quantitative agreement with the experimental result shown in Fig.\,\ref{FigPrinciple}a.

Displayed in Fig.\,\ref{FigPrinciple}b, are the same resistive curves plotted as $G$$=$$1/R_{\rs{\square}}$ vs. temperature. Fitting them by the standard BKT critical formula  $G$$=$$G_0\exp(-{\rm const}/\sqrt{T/T_{{\mathrm{dec}}}-1})$, yields indeed $T_{\mathrm{dec}}$$\approx$\,$400$\,mK. Deviations from the BKT criticality at lowest temperatures in Fig.\,\ref{FigPrinciple}b,c
indicate the point where the typical size of the dissolving dipole matches the size of the finite electrostatic screening length $\lambda_{\mathrm c}$ which appears in the normal insulator due to screening effect mediated by the free charges (at the BKT transition $\lambda_{\mathrm c}\to\infty$, and as a practical matter, the perfect criticality would have been observed for $\lambda_{\mathrm c}>10^5$\,\cite{Mironov2018}). Following\,\cite{Mironov2018} and recalling that $G\propto n_f$, where $n_f$ is the density of free charge carriers, one obtains $\lambda_{\mathrm c}$ using it as an adjusting parameter to fit the curves in\,Fig.\,\ref{FigPrinciple}c. When doing that, one has to account for the image forces in the Poisson equation that lead to $\lambda_{\mathrm c}=\beta_{\rs L}L$, where $\beta_{\rs L}$ is a numerical coefficient that serves as an adjusting parameter and $L$ is the length of the bridge. The result in the inset in Fig.\,\ref{FigPrinciple}c perfectly confirms that $\lambda_{\mathrm c}\propto L$.

\section*{Comparison of experimental data with theoretical predictions}~~\\
Figure\,\ref{FigScurve} presents the threshold behaviors of the $I$-$V$ curves corresponding to bridges of different length for two samples having the resistances $R_{\rs \square}$\,$=$\,0.2 and 2\,M$\Omega$, respectively at $T=2$\,K. The current jumps span a range of a few orders of magnitude and become less sharp upon shortening the distance between electrodes. The threshold voltage $V_{\mathrm c2}$ exhibits a linear dependence upon $L$, see in Figs.\,\ref{FigScurve}c,d, implying that the threshold electric field $E_{\mathrm c2} = V_{\mathrm c2}/L$ is independent on the distance between electrodes in full accordance with (\ref{VvsL}). For the low-resistance sample, shown in Fig.\,\ref{FigScurve}a,
the string tension can be estimated from $d_s \simeq 0.13$mm as $\sigma \simeq 4.15 \times 10^{-21}$J/m = 26 meV/m, which leads to a contribution $\sigma /2e = 1.3 \times 10^{-5}$V/mm to the slope. This is negligible with respect to the Coulomb contribution (second term in (\ref{VvsL})), which amounts to a predicted slope of 0.63 V/mm, again in remarkable quantitative agreement with the measured slope of 0.78 V/mm. The $I$-$V$ characteristics below the threshold, measured at 50 mK, are linear, which agrees reasonably well  with the theoretically predicted power-law exponent $1+(T_{\rm dec}/16 T)= 1+(400{\rm mK}/ 16\times 50{\rm mK}) = 1.5$.

The $I$-$V$ characteristics of the high-resistance sample are shown in Fig.\,\ref{FigScurve}b. The three largest bridges exhibit two-kink $I$-$V$ curves. The low-voltage kinks are identified as $V_{\rm c1}$, and the upper ones represent $V_{\rm c2}$. The two smallest bridges do not resolve $V_{\rm c1}$. This is because their size is so small that the strings penetrate the entire sample in the whole experimentally accessible voltage range and, thus, the Meissner state is not visible. 
The linear dependence $V_{\mathrm c2}\propto L$ of the bridge length $L$ has the slope 2.1\,V/mm, see Fig.\,\ref{FigScurve}d. Three available values of $V_{\rm c1}$ are apparently not sufficient for conclusive evidence of linearity, see Fig.\,\ref{FigScurve}e. Let us however \textit{assume} that $V_{\rm c1}\propto L$ in accordance with (\ref{exp}). Then the mean square deviation estimate for the approximating straight line (shown in red) would give 1.02\,V/mm for the slope. Making use Eq.\,(\ref{VvsL}), one would obtain the slope $2.1$$-$$1.02$$\approx$$1.1$ characterizing the pure Coulomb contribution. From (\ref{VvsL}), we then estimate the deconfinement temperature 692\,mK for this sample, yielding the exponent $1+(T_{\rm dec}/16 T)= 1+(692{\rm mK}/ 16\times 50{\rm mK}) = 1.86$ for the $I$-$V$ characteristics in the mixed state, $V_{\rm c1} < V < V_{\rm c2}$. Comparing this value with the measured ones for the three largest bridges, 1.6, 1.8 and 2, respectively, we conclude that the assumption $V_{\rm c1}\propto L$ results in a fair quantitative agreement between the predicted and measured exponents. Thus, the observation of $V_{\rm c1}$ is the first ever direct measurement of  Polyakov's string tension.

Finally let us discuss why the kink associated with $V_{\mathrm{c1}}$ is not seen in the IV-curves of the low-resistance sample. Due to its lower resistance, this sample is closer to the SIT, where the deconfinement temperature decreases towards zero. Since measurements of both samples are made at the same temperature, this means that the low-resistance sample is closer to the deconfinement temperature, where the string tension vanishes and strings become infinitely long. As a consequence, strings in the lower-resistance sample are longer than in the high-resistance sample and penetrate the sample end-to-end for all accessible voltages, so that the lower kink $V_{\rm c1}$ is not observable. In order to resolve the lower kink and access the Meissner state one needs larger samples, which are sufficiently far from the the SIT and have measurements carried out at low enough temperatures so that the kink was not fogged by the noise. 

\section*{A theory of the electric Meissner state}~~\\
Now we discuss the electric Meissner effect and formation of the electric Meissner state at electric fields $E<E_{\mathrm{c1}}$ which is another spectacular manifestation of Cooper pair confinement.
The action of the Abelian confining string, in its local formulation, is induced by the antisymmetric tensor gauge field of the second kind\,\cite{conf, dqt}. The vector current of point-like particles is replaced by the tensor current ${\cal J}^{\mu \nu}$ of strings (Greek letters denote space-time indices, Latin letters stand for spatial indices, and we use natural units $c$$=$$\hbar$$=$$1$).  Accordingly, the vector potential $A_{\mu}$, related to the electromagnetic field via $F_{\mu \nu} = \partial_{\mu} A_{\nu} -\partial_{\nu} A_{\mu}$, and coupling to the particle vector current, is replaced by the fundamental field tensor ${\cal F}_{\mu \nu}$ coupled directly to the string tensor current ${\cal J}^{\mu \nu}$. The equations for this field\, see Methods
\begin{equation}
\left( \Box + m^2 \right) {\cal F}^{\mu \nu} = 2\Lambda^2 {\cal J}^{\mu \nu}  \ ,
\label{wave}
\end{equation}
\begin{equation}
\partial_{\alpha} \left( \partial_{\mu} {\cal F}^{\mu \nu} \right) - \partial_{\nu} \left( \partial_{\mu} {\cal F}^{\mu \alpha} \right) = 0 \ ,
\label{gauge}
\end{equation}
with $m=\Lambda/e$, $\Lambda=\Lambda_0 \sqrt{z}/4$, can be viewed as the dual London equations. 
In a superinsulator with the light speed $v$ they become\,see Methods.
\begin{equation}
\left( \Box + (mv^2)^2 \right) {\cal F} ^{\mu \nu} = 2 \left(\Lambda v^2 \right)^2 \tilde {\cal J}^{\mu \nu}\,,\,\,
\Box\equiv\partial_0 \partial_0 -v^2 \nabla^2 \,,
\label{wavev}
\end{equation}
where $\tilde {\cal J}^{0i} = v{\cal J}^{0i}$, $\tilde {\cal J}^{ij} = (1/v){\cal J}^{ij}$, and time derivatives and factors ${\cal F}^{0i}$ in (\ref{gauge}) must be substituted by $(1/v)\partial_0$ and $(1/v)\  {\cal F}^{0i}$, respectively.
Equation\,(\ref{wavev}) implies the electric Meissner effect for applied voltages below a critical value $V_{\rm c1}$ where strings do not penetrate the sample. In the static situation with no strings, Eq.\,(\ref{wavev}) reduces to
\begin{equation}
\left( \nabla^2- (mv)^2 \right) {\cal F}^{0i} = 0 \ .
\label{poisson}
\end{equation}
For a superinsulator occupying the $z>0$ half-space and a uniform electric field $E_{\rm ext}$ applied in the $x$ direction with boundary condition ${\cal F}^{01} (z=0) = E_{\rm ext}$ one finds
\begin{equation}
{\cal F}^{01} (z) = E_{\rm ext} \exp(-z/\lambda_{\mathrm{el}})\,,\,\,\,\,\,\,z>0\,,
\label{solution}
\end{equation}
where $\lambda_{\rm el}$$=$$1/(vm)$ is the electric analogue of the London penetration length. Hence, in a bulk superinsulator in the electric Meissner state, the static dielectric permittivity $\varepsilon_{\rs SI}$$=$$0$.

To describe the electric Meissner effect in terms of magnetic monopoles, let us recall that the electrodynamics of the dual charge-monopole ensemble is governed by the Maxwell-Dirac equations\,\cite{goddard}
\begin{equation}
\partial_{\mu } F^{\mu \nu}= j_q^{\nu} \ ,\,\,\, \partial_{\mu } \tilde F^{\mu \nu}= j_{\phi} ^{\nu} \ ,
\label{Dirac}
\end{equation}
where $\tilde F^{\mu \nu} = (1/2)\epsilon^{\mu \nu \alpha \beta} F_{\alpha \beta} $ is the dual electromagnetic field tensor ($\epsilon^{\mu \nu \alpha \beta} $ is the totally antisymmetric tensor), $j_q^{\mu}$ is the charge current and $j_\phi^{\mu}$ is the magnetic monopole current.  Consider an infinite superinsulating slab harboring an ensemble of vortices aligned along the $z$ axis perpendicular to the slab surfaces, with their magnetic monopole endpoints.  The applied dc electric field $\mathbf{E}\equiv(E_x,0,0)$ generates a magnetic monopole current, according to the dual Amp\`ere law $\nabla \times {\bf E} = - {\bf j}_{\phi}$, circulating around the slab in the $\pm y$-direction, which, in turn, induces a shielding electric field opposite to the applied one. Since, in the condensate, monopole motion is dissipationless, the entire external applied electric field is screened. Hence the electric Meissner effect.

The above consideration captures the physics of bulk superinsulators, but the case of thin films is more subtle. 
The screening length $\lambda_{\mathrm{el}}\lesssim  d_{\mathrm s} \simeq \sqrt{v/\sigma}$\,\cite{dtv1}, increases with temperature and diverges at the deconfinement transition\,\cite{dtv2}. For TiN, $d_{\mathrm s}\simeq 60\,\mu$m\,\cite{dtv1}. In typical experiments on TiN and NbTiN films, where the film thickness $d=3-20$\,nm\,$\ll\lambda_{\mathrm{el}}$, 
the concept of circular monopole currents does not apply literally, since the physical size of the ``bulk" monopoles exceeds the film thickness. In this 2D case, monopoles become instantons\,\cite{polyakov}, describing tunneling events at which vortex ``particles" appear and disappear in the condensate. 

\section*{Discussion}~~\\
The novel material presented here, a superinsulator based on an atomic layer deposition grown NbTiN film, opens a new route for further exploring the superinsulating state. Mapping the superinsulator onto compact QED enabled us to construct an electric analogue of the superconducting London theory and to describe the electric response and the $I$-$V$ characteristics of a superinsulator. The device geometry, a Hall bar pattern, allowed us to investigate the details of the electric behavior of a superinsulator as function of the system size and to reveal the linear dimension dependencies of both critical voltages, $V_{\mathrm c1}$, at which the electric mixed state forms, and $V_{\mathrm c2}$, at which superinsulation breaks down. These observations open the door for a direct observation of asymptotic freedom phenomena and allow first-ever measurements of the linear tension of Polyakov's strings and investigations of the interior of an electric meson via desktop experiments.

Considering the crudeness of our long-wavelength model, the observed  $V_{\mathrm c2}(L)$ dependencies demonstrate an amazingly good agreement with the theoretical predictions. At the same time, while the results for  $V_{\mathrm c1}$ are in an excellent concert with the theoretical estimates, provided the linear dependence $V_{\mathrm c1}$ is assumed,  more experimental research is required to establish a calibrated tool for measurements of the string tension. 







\subsection{Acknowledgements}
	This work was supported by the U.S. Department of Energy, Office of Science, Basic Energy Sciences,
	Materials Sciences and Engineering Division (V.M.V.). Experimental work (S.V.P. and A.Yu.M.) were supported by Russian Science Foundation project No.~18-72-10056. M.C.D. thanks CERN for hospitality during completion the work.
	We are grateful to Dr.\,T.\,I.\,Baturina for valuable contribution at the initial stage of research of superinsulator in NbTiN.

\section{APPENDIX}

\section{Electric London equations}~~\\
The tensor current describing the motion of strings replaces the vector current encoding the motion of point particles. Correspondingly, the vector potential $A_{\mu}$ related to the original electromagnetic field strength via $F_{\mu \nu} = \partial_{\mu} A_{\nu} -\partial_{\nu} A_{\mu}$ is replaced by a fundamental tensor field ${\cal F}_{\mu \nu}$ coupled to the tensor current ${\cal J}^{\mu \nu}$.
The associated field strength is given by the three-tensor ${\cal H}_{\mu \nu \rho} = \partial_{\mu} {\cal F}_{\nu \rho} + \partial_{\nu} {\cal F}_{\rho \mu} + \partial_{\rho} {\cal F}_{\mu \nu}$.
The Euclidean action for electric and magnetic fields in the superinsulating vortex condensate\,\cite{dqt} acquires the form (we use natural units $c=1$, $\hbar = 1$)
\begin{equation}
S= \int d^4 x \ {1\over 12 \Lambda^2} {\cal H}_{\mu \nu \rho} {\cal H}_{\mu \nu \rho} + {1\over 4e^2} {\cal F}_{\mu \nu} {\cal F}_{\mu \nu}+ i {1\over 2} {\cal F}_{\mu \nu} {\cal J}_{\mu \nu} \ ,
\label{jt}
\end{equation}
where $\Lambda = \Lambda_0 \sqrt{z}/4$, with $\Lambda_0=1/r_0$ the UV cutoff and $z$ the magnetic monopole quantum fugacity.
Varying with respect ${\cal F}_{\mu \nu}$ yields the equations of motion
\begin{equation}
\partial_{\mu} {\cal H}^{\mu \alpha \beta} +m^2 {\cal F}^{\alpha \beta} = 2\Lambda^2 {\cal J}^{\alpha \beta} \ ,
\label{eqmotion}
\end{equation}
which, in turn reduce to Eqs.\,(11),(12) of the main text.

For the sake of transparency, we now show how they reduce to the usual Maxwell equations in the limit of vanishing monopole fugacity, $z\to 0$, when the vortex condensate disappears. In this limit we have  $\Lambda \to 0$ and finiteness of the partition function $Z = \int {\cal D} {\cal F}_{\mu \nu}\  {\rm exp} (-S)$ requires that the 3-tensor field strength vanishes, ${\cal H}_{\mu \nu \rho} = 0$. This is generically the case for ``pure gauge configurations" ${\cal F}_{\mu \nu} = \partial_{\mu} A_{\nu}-\partial_{\nu} A_{\mu}=F_{\mu \nu} $ so that, in this limit, the partition function of classical Maxwell electrodynamics is recovered,
\begin{equation}
Z \to \int {\cal D} A_{\mu }\  {\rm exp} \left( - \int d^4 x\  {1\over 4e^2} F_{\mu \nu} F_{\mu \nu} + i  A_{\mu} j_{\mu} \right) \ ,
\label{limit}
\end{equation}
with the current $j^{\mu} = \partial_{\nu} {\cal J}^{\mu \nu}$ describing the now free charges at the end of the original strings. This shows what is the relation between electrodynamics within the vortex condensate and the familiar Maxwell theory. The latter is recovered as a ``pure gauge" version of the former.
\\

\section{Electric field equations in materials with 
	$v=c$/$\sqrt{\varepsilon\mu}$}~~\\
The above ``electric London equations" hold if the speed of light is $c$ ($=1$ in our natural units), which is reflected in the Lorentz invariance of the action (\ref{jt}) and the Lorentz covariance of the equations of motion (11) of the main text. For real materials, however, this is surely not the case. The (Euclidean) action for electromagnetic fields in a linear material is given by
\begin{equation}
S_{\rm linear} = {1\over 2e^2} \int dt \ d^3 x \left( {1\over v}  {\bf E}^2 + v {\bf B}^2 \right) \ ,
\label{linear}
\end{equation}
where $v=1/\sqrt{\mu \varepsilon}$ is the light velocity in the material, expressed in terms of the dielectric permittivity $\varepsilon$ and the magnetic permeability $\mu$.  Due to $v<1$, the symmetry of this action is now restricted to non-relativistic Galilean invariance. Correspondingly, the Galilean-invariant action for a superinsulating thin film of this material when vortices condense is
\begin{eqnarray}
S = \int dt\  d^3 x \ {1\over 12 \ v^2 \Lambda^2} \left( v {\cal H}_{ijk}{\cal H}_{ijk} + {3\over v} {\cal H}_{0ij}{\cal H}_{0ij} \right)
\nonumber \\
+{1  \over 4e^2} \left(v {\cal F}_{ij} {\cal F}_{ij} +2{1\over v}
{\cal F}_{0i}{\cal F}_{0i} \right) + i {1 \over 2} {\cal F}_{\mu \nu} {\cal J}_{\mu \nu} \ ,
\label{nonrel}
\end{eqnarray}
where $v$ is the light velocity expressed in terms of material parameters of the normal insulating state.
The equations of motion (11) and (12) are modified to
\begin{eqnarray}
&&\left( \Box + (mv^2)^2 \right) {\cal F} ^{\mu \nu} = 2 \left(\Lambda v^2 \right)^2 \tilde {\cal J}^{\mu \nu}
\nonumber \\
&&\Box = \partial_0 \partial_0 -v^2 \nabla^2 \ .
\label{wavenonrel}
\end{eqnarray}
where $\Delta_{\rm SI} = mv^2 $ represents the energy gap of the superinsulator, $\tilde {\cal J}^{0i} = v{\cal J}^{0i}$, $\tilde {\cal J}^{ij} = (1/v){\cal J}^{ij}$, and every time derivative and factor ${\cal F}^{0i}$ in the gauge condition, Eq.\,(12) of the main text,  must be substituted by $(1/v) \ \partial_0$ and $(1/v)\  {\cal F}^{0i}$, respectively.
\\
\section{Sample preparation}~~\\
The sample growths were carried out in a custom-made viscous flow ALD reactor in the self limiting regime. A constant flow of ultrahigh-purity nitrogen (UHP, 99.999\%, Airgas) at $\sim 350$ sccm with a pressure of $\sim 1.1$ Torr was maintained by mass flow controllers. An inert gas purifier (Entegris GateKeeper) was used to further purify the N$_2$ gas by reducing the contamination level of H$_2$, CO, and CO$_2$ to less than 1 ppb and O$_2$ and H$_2$O to less than 100 ppt. The thermal ALD growth of the AlN/NbTiN multilayer was performed using alternating exposures to the following gaseous reactants with the corresponding timing sequence (exposure-purge) in seconds: AlCl$_3$ (anhydrous, 99.999\%, Sigma-Aldrich) (1 - 10), NbCl$_5$ (anhydrous, 99.995\%, Sigma-Aldrich) (1 - 10), TiCl$_4$ (99.995\%, Sigma-Aldrich) (0.5 - 10) and NH$_3$ (anhydrous, 99.9995\%, Sigma-Aldrich) (1.5 - 10). The intrinsic silicon substrates were initially cleaned insitu using a 60 s exposure to 
$7.5\pm 0.5$ nm was deposited at 
450$^\circ C$ with 200 ALD cycles. The chamber 
temperature was then lowered to 350$^\circ C$ to synthesize the NbTiN layers. The growth cycle of the NbTiN is $2\times$(TiCl$_4$ + NH$_3$) 
and $1\times$(NbCl$_5$ + NH$_3$) that was repeated 80 times with the corresponding total ALD cycles 240 to produce the film thicknesses 10 nm as measured ex-istu by X-ray reflectivity (XRR). The chemical composition measured by X-ray Photoemission Spectroscopy (XPS) and Rutherford Backscattering Spectroscopy (RBS) show consistently for the AlN layer $5.5\pm0.3$\% of Cl impurities and a Al/N ratio of $1\pm0.05$, whereas for the NbTiN films $3\pm0.3$\% of Cl impurities, a Nb/Ti ratio of $2.3\pm0.03$ and a (Nb+Ti)/N ratio of $1\pm0.03$. The material densities measured by RBS and XRR are $2.5\pm0.01$ g/cm$^3$ in AlN and $6\pm0.05$ g/cm$^3$ in the NbTiN. After deposition films were stored at room conditions in ten years.

\section{Samples parameters and measurements}~~\\
The sample parameters are as follows: the diffusion constant $D \sim 0.25 cm^2s^{-1}$, the superconducting coherence length $\xi(0)\sim 5$ nm, the density of states as $n\sim 4\cdot10^{21} cm^{-3}$~\cite{Hall}.
NbTiN film was patterned using photolithography and plasma etching into resistivity bar 50 $\mu$m wide and with 100, 250, and 100 $\mu$m separation between additional perpendicular bars. On this additional bars were making gold contacts. The chosen design allows for two-probe resistivity measurements of regions with different length (2.4, 1.34, 0.55, 0.44, 0.32, 0.2 $\mu$m). Measurements of the temperature  dependencies of the resistance were carried out in $^{3}$He/$^{4}$He dilution refrigerator Triton400. Two-probe technique was used with the 100 $\mu$V and 1 Hz. The chosen values of applied voltage guaranteed that the resistance was obtained in the linear response regime by the direct measurements of current-voltage characteristics. Ac measurements were performed using SR830 lock-in amplifier, dc measurements were performed using nanovoltmeter Agilent 34420, current was transformed into voltage with using SR570 low-noise current preamplifier with filtration system.
\\
\section*{Data availability}~~\\
The authors declare that all relevant data supporting the findings of this study are available
within the article.


\hskip 1pt

\end{document}